# SSA-Caterpillar in Group Anonymity

Dan Tavrov
*Applied Mathematics Department*
*National Technical University of Ukraine*
*"Kyiv Polytechnic Institute"*
*Kyiv, Ukraine*
dan.tavrov@i.ua

Oleg Chertov
*Applied Mathematics Department*
*National Technical University of Ukraine*
*"Kyiv Polytechnic Institute"*
*Kyiv, Ukraine*
chertov@i.ua

*Abstract* – Nowadays, it is a common practice to protect various types of statistical data before publishing them for different researches. For instance, when conducting extensive demographic surveys such as national census, the collected data should be at least depersonalized to guarantee proper level of privacy preservation. In practice, even more complicated methods of data protection need to be used.

All these methods can be generally divided into two classes. The first ones aim at providing individual data anonymity, whereas the other ones are focused on protecting information about a group of respondents.

In this paper, we propose a novel technique of providing group anonymity in statistical data using singular spectrum analysis (SSA). Also, we apply SSA to defining hidden patterns in demographic data distribution.

## I. INTRODUCTION

"He who controls the information, controls the world." This famous quote vividly expresses one of the main features of a post-industrial world. Today, necessity of collecting, storing, and especially processing different kinds of information is understood well by those who want to keep up with the pace of contemporary society.

Without a doubt, national population census is the most extensive activity of collecting large amounts of data. The aggregated census data often are accessible in suitable forms like statistical tables or graphs, and on suitable convenient media (paper publications or OLAP databases). Also, they are disseminated by means of suitable channels such as publishers and the Internet [1]. At the same time, big samples of the primary census data usually are reduced to one *microfile* containing gathered information about *respondents*, possibly, in a coded form (see Table 1). Obviously, necessary precautions should be taken beforehand in order to guarantee acceptable level of data privacy.

TABLE I
MICROFILE DATA IN MATRIX FORM

|  |  | Attributes | | | |
|---|---|---|---|---|---|
|  |  | $u_1$ | $u_2$ | ... | $u_\eta$ |
| Respondents | $r_1$ | $\omega_{11}$ | $\omega_{12}$ | ... | $\omega_{1\eta}$ |
|  | $r_2$ | $\omega_{21}$ | $\omega_{22}$ | ... | $\omega_{2\eta}$ |
|  | ... | ... | ... | ... | ... |
|  | $r_\mu$ | $\omega_{\mu 1}$ | $\omega_{\mu 2}$ | ... | $\omega_{\mu\eta}$ |

By data privacy we understand a user protection against discovery and further misuse of identity by other users [2, p. 69]. According to ISO/IEC 15408-2:2008, privacy consists of four family functional requirements, anonymity being the principal one.

In the following subsection, we will cover some approaches to providing anonymity in microfiles.

### A. Anonymity

In a consolidated proposal for terminology [3], anonymity of a subject means that the subject is not identifiable (uniquely characterized) within a set of subjects. So, in this work, we will treat both individual and group anonymity as a particular subject's property of being unidentifiable within a microfile. But, as individual anonymity concerns individual microfile records (such as citizens, households, and enterprises), group anonymity has to do mainly with certain sets of records defined in a specific way (to be discussed later on in this paper).

All the methods of providing individual anonymity are usually referred to as part of privacy preserving data publishing (PPDP) methods. The name clearly states that some kind of modification should be applied to the data before making them widely accessed. A state-of-the-art survey of PPDP is given in [4]. Here, we will only outline the most frequently used individual anonymity methods.

So-called *perturbative* PPDP methods achieve data anonymity by altering dataset records in a particular way. For instance, *randomization* [5] implies adding noise to initial data so that it is hard to identify records with outstanding attribute values. Another approach lies in gaining *k-anonymity* [6], which means that at least *k* dataset records correspond to a specific attribute values' combination. And, of course, it is possible to *swap* attribute values between data records [7].

Among recently developed perturbative PPDP methods we could mention using Fourier transform [8] and singular value decomposition [9].

All the methods mentioned earlier are mainly used in *non-interactive* privacy mechanisms when the trusted data collector publishes a cleansed version of the collected data. The other situation arises when applying *interactive* privacy mechanisms when the data collector provides means for the users through which they can make queries and obtain answers about the data. The latter case is somewhat easier to deal with, mainly due to the fact that the data collector already knows what kind of utility the user wants to receive. Nevertheless, there are some specific problems to face (for instance, even if information on a particular individual isn't present in the dataset at all, privacy breach can occur provided that additional sources of information are available). To cope with such issues, the concept of *differential privacy* has been introduced and successfully developed (e.g., refer to [10]).

Along with perturbative methods, one might also apply various *non-perturbative* methods such as *data recoding* and *data suppression* [11]. These methods, unlike perturbative ones, anonymize the data without altering them.

As opposed to PPDP methods, those ones for providing group anonymity should be capable of modifying not only particular respondents' attribute values but their distribution as well. Moreover, they have to ensure that data distortion being introduced is satisfactory. That is why completely new techniques need to be developed.

At the time this paper is being written, maybe the most complete source of how to provide group anonymity is [12]. The paper proposes a generic scheme of providing group anonymity and illustrates it with some practical examples.

One of the ways to provide group anonymity described in [12] is based on *normalization* process presented in [13]. This makes it possible to preserve major data statistical features, e.g. mean value and standard deviation. Though, it does not seem to be quite sufficient.

Another approach is based on applying wavelet transform (WT) [14] to initial data so that it helps to conceal important data features not likely to be given away. At the same time, it enables preserving data frequency peculiarities. But, this technique has some slight disadvantages, some of which are:

*1) Insufficient Utility Preserving*: The way utility is being preserved isn't clear enough.

*2) Strong Dependence on the Algorithm*: Choosing different wavelet bases yields completely different outcomes. At the same time, there are no well-defined means of defining which base should be chosen.

*3) Tight Constraints while Modifying the Data*: When performing data perturbation, the researcher is highly forced to modify dataset records in a way implied by the method itself.

All these disadvantages lead to a conclusion that using WT for providing group anonymity is a very much method-oriented technique. Fortunately, there exists quiet dissimilar approach which is free of all three downsides mentioned above. Indeed, applying SSA, as we propose it in this paper, gives a researcher powerful means of modifying initial data almost deliberately and simultaneous preserving data utility.

*B. Singular Spectrum Analysis*

Singular spectrum analysis (SSA) (traditionally called "Caterpillar" in papers written by Russian researchers) is a relatively novel technique for analyzing time series of quite dissimilar origin. SSA-Caterpillar method finds its applications in fields ranging from climate research [15] and geophysics [16] to processing time series with missing values [17] and detecting spatial data patterns [18].

Generally speaking [17], the method's aim is to split original time series into a sum of trend, periodic, and/or noise components. What makes this technique especially powerful is that the original series do not have to be stationary, the model of the trend can be completely indeterminate, and we do not even have to know anything about periodic components within the series.

Due to main peculiarities of the method, it cannot be fully automated. This means that the last decision is always left to the researcher. Moreover, principle parameters of the method should be also picked by a human in each and every case independently. This leads to some complications in applying the method to analyzing large amounts of information. Though, despite this fact, it is mostly efficient in performing such tasks as series smoothing, predicting, and filling in missing signal values.

SSA-Caterpillar technique was independently developed in the UK and the USA, from one side, and Russia, from the other one. At present, there exist rather thorough and complete studies of the method full of practical examples, among which [19] and [20] can be mentioned.

## II. THEORETIC BACKGROUND

*A. Group Anonymity*

In this subsection, we will present main steps of providing group anonymity in microfiles, though we will omit some facts not necessary for the further discussion. For more complete explanation and real-life illustrations, refer to [12].

Let us propose some necessary definitions.

*Definition 1. An identifier* is a microfile attribute which unambiguously defines a certain respondent in a microfile.

Since keeping identifiers in the microfile when publishing it is the best way to violate individual privacy, they should be always removed from the dataset. Further on, we will presume that any microfile does not contain any identifier.

*Definition 2*. We will call certain attributes $u_{v_j}$, $j = \overline{1,t}$ *vital attributes*. A subset of a Cartesian product of vital attributes $u_{v_1} \times u_{v_2} \times ... \times u_{v_t}$ will be called *a vital set*.

Vital attributes help define subsets of microfile respondents to be protected when providing group anonymity.

*Definition 3*. We will call an element $s_k^{(v)} \in S_v$, $k = \overline{1, l_v}$, $l_v \leq \mu$, *a vital value combination*. Each $s_k^{(v)}$ element will be called *a vital value*.

Specific vital value combinations distinguish those respondent categories whose distribution is supposed to be protected.

*Definition 4*. We will call an element $s_k^{(p)} \in S_p$, $k = \overline{1, l_p}$, $l_p \leq \mu$, $S_p$ being a subset of microfile data elements corresponding to the $p^{th}$ attribute (called *a parameter set*), *a parameter value*. The attribute itself will be called *a parameter attribute*.

We usually use parameter values to arrange microfile data in some particular order, so that it becomes possible to treat them as certain series.

*Definition 5.* A *group* $G(V,P)$ is an attribute set consisting of several vital attributes $V = \{V_1, V_2, ..., V_l\}$ and a parameter attribute $P$, $P \neq V_j$, $j = 1,...,l$.

So, providing group anonymity actually lies in performing such microfile modifications (with respect to each group $G_i(V_i, P_i)$, $i = 1,...,k$ separately) that sensitive data become completely confided. Or, speaking more formally, we can divide providing group anonymity into the following stages:

*1) Data Preparation:* Construct a (depersonalized) microfile **M** representing statistical data to be processed.

*2) Defining Goals and Targets:* Define one or several groups $G_i(V_i, P_i)$, $i = 1,...,k$ standing for respondents categories to be protected.

*3) Group Anonymity Calculations:* For each $i$ from 1 to $k$:

*a) Choosing data representation*: Pick a proper data representation (called *goal representation*) $\Omega_i$ (**M**, $G_i$) for a group $G_i(V_i, P_i)$.

*b) Performing data mapping*: Define a mapping function
$$\Upsilon : \mathbf{M} \to \Omega_i (\mathbf{M}, G_i) \quad (1)$$
(called *goal mapping function*). Obtain needed goal representation (see the previous step) of a dataset.

*c) Performing goal representation's modification*: Define a functional
$$\Xi : \Omega_i (\mathbf{M}, G_i) \to \Omega'_i (\mathbf{M}, G_i) \quad (2)$$
(also called *modifying functional*). Obtain a modified goal representation.

*d) Obtaining the modified microfile.* Define an *inverse goal mapping function*
$$\Upsilon^{-1} : \Omega'_i (\mathbf{M}, G_i) \to \mathbf{M}^*. \quad (3)$$
Obtain a modified microfile.

*4) Post-Processing:* Prepare the modified microfile for publishing.

We introduced some new terms in the scheme above, so we would like to clarify them a bit before moving forward.

*Definition 6.* A *goal representation* $\Omega$ (**M**, $G$) of a dataset **M** with respect to a group $G$ is a dataset reflecting specific features of a group within the microfile in a way appropriate for providing group anonymity.

In general, this dataset could be of any possible form convenient for the researcher. In this paper, we will work with one particular goal representation called *a quantity signal* $q = (q_1, q_2, ..., q_m)$. It can be constructed by counting up all the respondents in a group with a certain pair of vital value combination and a parameter value, and arranging them afterwards in any order proper for a parameter attribute.

A quantity signal provides a quantitative statistical distribution of group members from initial microfile. In some situations, extremums of this distribution could lead to disclosing restricted information, which makes it a suitable goal representation.

*Definition 7.* A *modifying functional* $\Xi : \Omega_i$ (**M**, $G_i$) $\to$ $\to \Omega'_i$ (**M**, $G_i$) of a dataset **M** with respect to a group $G$ is any function, algorithm, or procedure which transforms initial goal representation into another one ensuring that group anonymity is reached.

Frankly saying, creating modifying functional is possibly the toughest, though yet utterly important part of the whole providing group anonymity process. In this paper, we will pay most attention to one version of modifying functional based on applying SSA-Caterpillar technique.

*B. SSA-Caterpillar*

In the current subsection, we will discuss main steps of the classical SSA-Caterpillar algorithm. Also, we will propose some pieces of advice on how to correctly pick method's parameters to receive needed outcome. We will heavily base our explanation on [21].

So, let us be given initial non-zero time series $F = (f_0, ..., f_{N-1})$. Then, to split the series into trend, periodic, and noise components, we need to carry out following steps:

*1) Embedding:* Transform the series into *a* so-called *trajectory matrix*:

$$\mathbf{X} = \begin{pmatrix} f_0 & f_1 & f_2 & \cdots & f_{K-1} \\ f_1 & f_2 & f_3 & \cdots & f_K \\ f_2 & f_3 & f_4 & \cdots & f_{K+1} \\ \vdots & \vdots & \vdots & \ddots & \vdots \\ f_{L-1} & f_L & f_{L+1} & \cdots & f_{N-1} \end{pmatrix}. \quad (4)$$

In (4), $L$ stands for the only method's parameter at this stage called *window length*, $1 < L < N$, $K$ stands for the number of *lagged vectors* obtained as a result of embedding procedure, $K = N - L + 1$.

Trajectory matrix is a Hankel matrix, i.e. all the elements along the $i + j = const$ diagonal are the same.

*2) Singular Value Decomposition:* Present (4) as a sum of rank-one bi-orthogonal elementary matrices.

To do that, let's define $\mathbf{S} = \mathbf{X}\mathbf{X}^T$. Then, by $\lambda_1, ..., \lambda_L$ we will understand *eigenvalues* of **S** arranged in non-increasing order $\lambda_1 \geq ... \geq \lambda_L \geq 0$, whereas $U_1, ..., U_L$ will denote orthonormal system of **S** eigenvectors corresponding to these eigenvalues.

Let $d = \max\{i : \lambda_i > 0\}$. Then, if to denote $V_i = \mathbf{X}^T U_i / \sqrt{\lambda_i}$, $i = \overline{1, d}$, we can present decomposition of (4) the following way:
$$\mathbf{X} = \mathbf{X}_1 + ... + \mathbf{X}_d. \quad (5)$$
In (5), $\mathbf{X}_i = \sqrt{\lambda_i} U_i V_i^T$.

Singular value decomposition (5) is optimal in the sense that it provides the best approximation of (4) among all the other matrices of rank lower than *d*.

*3) Grouping:* Divide the index set $\{1,...,d\}$ into $m$ disjoint subsets $I_1,...,I_m$.

Then, the *resulting matrix* $\mathbf{X}_I$ corresponding to subset $I = \{i_1,...,i_p\}$ can be defined as

$$\mathbf{X}_I = \mathbf{X}_{i_1} + ... + \mathbf{X}_{i_p}. \quad (6)$$

From (6), it is evident that the original trajectory matrix (4) can now be viewed as the following sum:

$$\mathbf{X} = \mathbf{X}_{I_1} + ... + \mathbf{X}_{I_m}. \quad (7)$$

The last thing to complete is to return back from the matrix representation of our data to the series one.

*4) Diagonal Averaging:* Transfer each resulting matrix (6) into a time series.

To carry this operation out, let us first introduce the following notation: $L^* = \min(L,K)$, $K^* = \max(L,K)$, $y^*_{ij} = y_{ij}$, $L < K$, and $y^*_{ij} = y_{ji}$, $K < L$. Then, we can obtain time series $G = (g_0,...,g_{N-1})$ the following way:

$$g_k = \begin{cases} \dfrac{1}{k+1}\sum_{m=1}^{k+1} y^*_{m,k-m+2}, & 0 \le k < L^* - 1 \\ \dfrac{1}{L^*}\sum_{m=1}^{L^*} y^*_{m,k-m+2}, & L^* - 1 \le k < K^* \\ \dfrac{1}{N-k}\sum_{m=k-K^*+2}^{N-K^*+1} y^*_{m,k-m+2}, & K^* \le k < N \end{cases} \quad (8)$$

Finally, applying (8) to resulting matrices from (7), we obtain time series representation of our initial dataset:

$$F = \sum_{k=1}^{m} F^{(k)}. \quad (9)$$

In (9), $F^{(k)}$ stands for the time series obtained after diagonal averaging of the resulting matrix $\mathbf{X}_{I_k}$.

As we can see, the only two stages which can be completely left to the computer are the second and the fourth. It is the core feature of SSA-Caterpillar algorithm that the window length $L$ and exact list of index subsets $I_1,...,I_m$ should be picked by the researcher all on his own.

Though, there are some general hints to aid in choosing proper values:

*1)* If the time series $F^{(k)}$ corresponds to the resulting matrix $\mathbf{X}_{I_k}$, where $I_k$ consists of only one element, then its form will strongly resemble the one of the corresponding eigenvector. This means that if any eigenvector is periodic than the corresponding time series might be periodic as well. Moreover, the series trend is highly supposed to correspond to the eigenvector with relatively low frequency.

*2)* Two rather close singular values could be the sign of a periodic time series component. This means that we can extract periodic components by grouping resulting matrices corresponding to such pairs of singular values.

*3)* The greater is the eigenvalue, the greater is the contribution of the corresponding time series component to the time series in general.

*4)* Relatively small eigenvalue tending to zero are likely to point at noise time series components.

*5)* There is no need in choosing window length $L$ greater than $N/2$. This is true because singular value decompositions of trajectory matrices obtained with window lengths $L$ and $N - L + 1$ are equivalent.

*6)* In most cases, the greater is $L$, the more accurate is time series decomposition. That's why it seems reasonable to take $L$ as big as possible (and, according to the previous hint, the most appealing option is $N/2$). Though, in some particular cases like extracting periodic component it is advisable to pick window length being multiple of the component's period.

*C. Applying SSA-Caterpillar to Group Anonymity*

In this subsection, we will discuss how exactly SSA-Caterpillar technique might be applied to providing group anonymity in a microfile.

According to the group anonymity scheme presented at the beginning of this Section, the only possible stage to make use of SSA-Caterpillar is constructing the modifying functional (2). But, first we need to define what restrictions does this functional have to satisfy.

On one hand, the modifying functional has to be efficient for altering the quantity signal obtained from the original dataset. In this paper, we propose to alter the quantity signal by replacing its trend component from (9) with another one ensuring that original extremums of the signal are changed. In this case, we are able to conceal any extreme quantities from vital values' distribution which might occur to be security-intensive.

On the other hand, it is vitally necessary to make sure that data utility isn't reduced much. It is obvious that SSA-Caterpillar method is especially convenient for extracting periodic components from the original signal. Such components may be very important for statistical, demographic, and sociological researches (and this is why microfile data are used for in the first place).

So, we come to conclusion, that modifying trend component of the quantity signal with simultaneous preserving all the other components (including periodic ones) seems to be a rather adequate way of providing group anonymity in a microfile. In the next Section, we will illustrate this concept with a practical example based on real-life data.

### III. PRACTICAL EXAMPLE

According to the group anonymity scheme from Section II, the first step we have to take is to build up a microfile with the data needed to be protected. We decided to work with the 5-Percent Public Use Microdata Sample File provided by the U.S. Census Bureau [22] concerning the 2000 U.S. census of population and housing microfile data in California.

The second step requires defining groups to be protected. For simplicity of calculations, we took only one group corresponding to the military personnel distributed by their age.

As it was stated before, in this paper we will use quantity signals as a goal representation of the microfile data. To construct one for our current example, we counted up all the people on active military duty and grouped them by their age, resulting in the following signal:

$q$=(14, 203, 589, 713, 675, 604, 498, 374, 299, 274, 231, 231, 191, 212, 176, 175, 158, 185, 159, 185, 173, 145, 119, 118, 101, 74, 72, 54, 42, 32, 40, 30, 19, 13, 15, 10, 13, 5, 3, 3).

It is important to note here that due to the nature of the group under protection, we limited our parameter set down to values from 17 to 56 years of age. The graphical representation of the signal can be found in Fig. 4a.

As we can clearly notice, there are huge extremums at the $3^{rd}$, the $4^{th}$, the $5^{th}$, and the $6^{th}$ signal positions. To provide group anonymity, we need to replace the quantity signal's trend with another one, so that these extremums are not that outstanding.

In general, we could try to transfer these extremums to some other signal elements, or make an attempt at creating new alleged maximums. But, both of these approaches either yield not very truthful data, or cannot be completed in terms of trend substituting. That is why our only option left is to somehow smooth the signal so that these extremums cannot be marked out.

At the same time, as we mentioned in the previous Section, we need to preserve periodic components of the signal to reduce loss of data utility. To be able to accomplish both tasks, we need to decompose the signal first.

For that matter, we decided to take the value of 20 as the window length (mainly because the length of the whole signal is 40). After performing quite obvious steps of constructing trajectory matrix $\mathbf{X}$ according to (4) and obtaining resulting matrices (5) (which we omit in this paper), we need to group the components to extract valuable periodic and trend ones.

To make correct choice, we need to follow the instructions given at the end of Subsection II-C, especially one saying about tracking down pairs of almost equal $\mathbf{XX}^T$ eigenvalues likely to correspond to periodic signal components. That is why we present the square roots of these eigenvalues in Fig. 1a.

Unfortunately, there isn't any pair of similar eigenvalues present there. This means there is no evident periodic component within the original quantity signal. This situation is not quite preferable, because in this case there is no need in using such powerful technique as SSA-Caterpillar only to substitute the signal's trend.

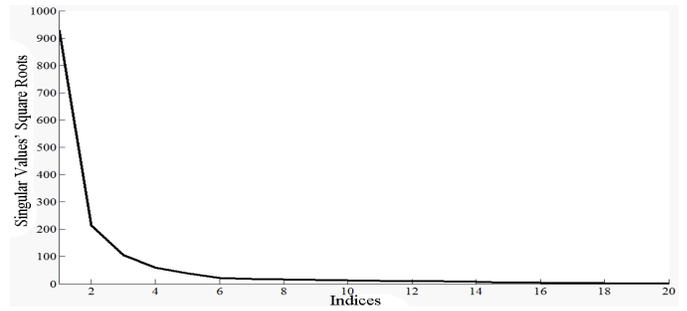

a)

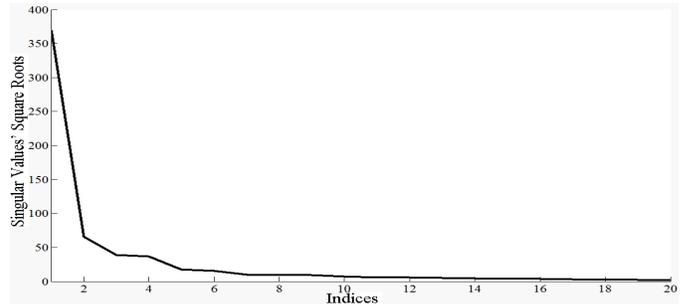

b)

Fig. 1 Square roots of eigenvalues of the trajectory matrix product by itself transposed. Cases with the quantity signals $q$ (a) and $\hat{q}$ (b).

Though, it seems to be not very believable that the time series representing military personnel of the state of California does not contain any periodic component at all. Indeed, if we take a closer look at the data we are working with, we will find out that the main reason of such an incident is that California is the home state for San Diego military bases being the largest concentration of naval facilities in the world. So, we are strongly convinced that periodic components, if there is any, are simply suppressed by the outweighing number of young military men working at San Diego naval bases.

To adjust our sampling procedure and make it more precise, we decided to exclude San Diego and nearby territories from our microfile. As a result, we obtain another quantity signal:

$\hat{q}$ =(2, 86, 223, 241, 227, 193, 152, 140, 95, 121, 92, 105, 87, 89, 79, 80, 83, 93, 78, 85, 79, 61, 62, 63, 59, 30, 38, 28, 24, 16, 21, 16, 10, 11, 4, 3, 7, 4, 2, 3).

This signal is present in Fig. 4b.

Once again, let us take a look at Fig. 1b representing $\mathbf{XX}^T$ eigenvalues (though, now the trajectory matrix corresponds to signal $\hat{q}$). In this case, we can mark out two clearly visible pairs of eigenvalues likely to correspond to the periodic components. The first one is values 3 and 4, and the second one is values 5 and 6.

Also, concluding from the magnitude of the first two eigenvalues, we can say that they correspond to the trend signal component. Taking this all into consideration, we can finally group all 20 signal components into the following index subsets: $I_1 = \{1,2\}$, $I_2 = \{3,4\}$, $I_3 = \{5,6\}$, $I_4 = \{7 \div 20\}$.

After having obtained resulting matrices (6) and performing diagonal averaging (8), we can present our quantity signal as a sum of the following components:
$$\hat{q} = \hat{q}^{(1)} + \hat{q}^{(2)} + \hat{q}^{(3)} + \hat{q}^{(4)}. \qquad (10)$$

In (10), $\hat{q}^{(1)}$ stands for signal's trend, $\hat{q}^{(2)}$ and $\hat{q}^{(3)}$ both reflect periodic components, and $\hat{q}^{(4)}$ will be treated as noise. Appropriate graphical representations can be seen in Fig. 2, and numerical values go as follows (we present here only three decimal numbers, though all the calculations had been carried out with much higher proximity):

$\hat{q}^{(1)}$ =(−10.345, 109.632, 205.501, 229.494, 215.163, 186.988, 159.323, 138.937, 123.873, 118.371, 111.498, 98.815, 92.343, 86.566, 82.496, 79.743, 76.639, 71.852, 67.303, 61.687, 56.087, 51.302, 47.627, 44.678, 42.162, 39.851, 37.710, 35.510, 33.610, 31.829, 30.166, 28.424, 26.547, 24.421, 22.043, 19.646, 17.442, 15.178, 13.169);

$\hat{q}^{(2)}$ =(−2.600, −3.897, 7.618, 11.991, 10.347, 4.727, −1.905, −7.448, −11.376, −12.155, −12.059, −10.619, −8.640, −5.751, −2.352, 1.611, 5.617, 9.017, 11.281, 12.633, 13.276, 12.816, 11.083, 8.460, 5.104, 1.165, −2.765, −6.736, −10.511, −13.715, −16.023, −17.522, −18.125, −17.906, −17.246, −16.284, −14.936, −13.391, −11.974, −10.411);

$\hat{q}^{(3)}$ =(14.835, −20.082, 11.729, −1.078, −1.038, 3.981, −7.909, 9.323, −11.634, 11.133, −8.375, 7.597, −6.236, 4.463, −4.432, 2.010, −2.120, 2.239, −1.246, 0.911, 0.713, −1.069, 2.129, −1.636, 2.793, −2.506, 1.975, −2.265, 0.674, −1.813, 0.676, −1.134, 0.490, −0.513, 0.530, −0.076, 0.209, 1.784, −0.839, 2.976);

$\hat{q}^{(4)}$ =(0.110, 0.347, −1.847, 0.593, 2.527, −2.696, 2.490, −0.812, −5.863, 3.650, 0.937, 2.166, 3.061, −2.055, −0.782, −6.116, −0.240, 5.105, −3.887, 4.153, 3.324, −6.833, −2.515, 8.549, 6.425, −10.822, −1.060, −0.709, −1.673, −2.082, 4.518, 4.490, −0.789, 2.873, −3.705, −2.684, 2.081, −1.835, −0.366, −2.735).

Judging from Fig. 2b and Fig. 2c, we can confidently admit that there are two periodic components in the quantity signal under analysis. The first one has a period of 20, and the second one seems to possess a period of 4. This is a very interesting result, though we are not qualified sociologists to make any proper conclusions about what these periods stand for. Nevertheless, such information has to be preserved for further studies.

To achieve group anonymity, we need to somehow smooth our trend component so that initial maximums are not so evidently observed. Surely, there might as well be other satisfactory trends to add. But, in this paper we would like to pick such a trend whose first 10 values are way too smaller than those of the existent one:

$\hat{q}^{(1')}$ =(−10.345, 96.508, 109.656, 117.067, 121.224, 123.347, 124.000, 123.347, 121.224, 117.067, 111.498, 105.856, 98.815, 92.343, 86.566, 82.496, 79.743, 76.639, 71.852, 67.303, 61.687, 56.087, 51.302, 47.627, 44.678, 42.162, 39.851, 37.710, 35.510, 33.610, 31.829, 30.166,

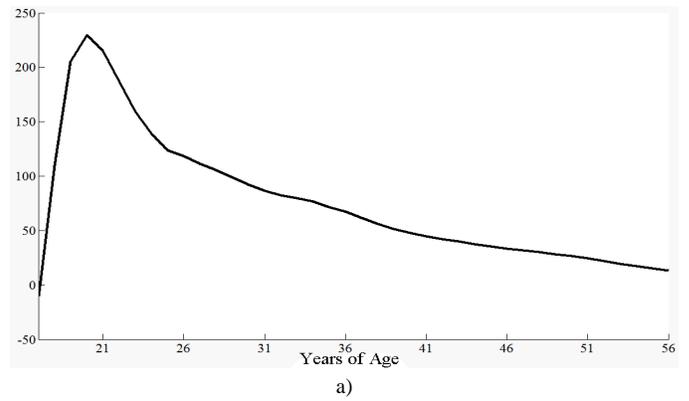

a)

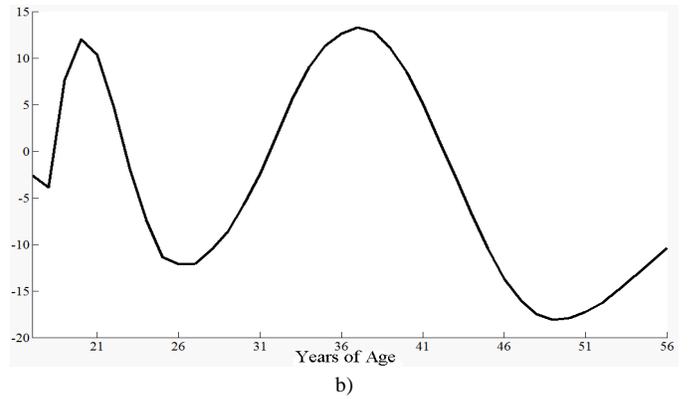

b)

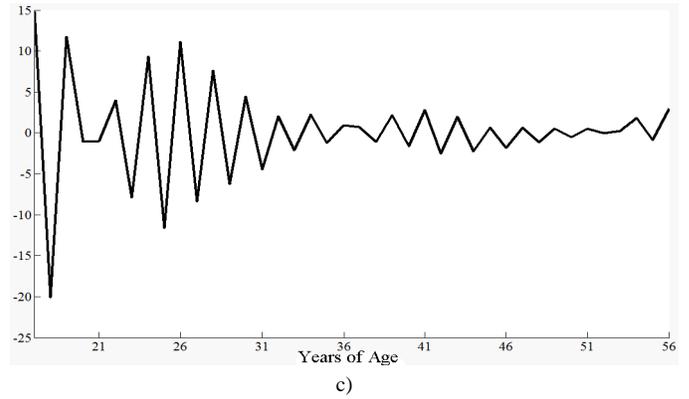

c)

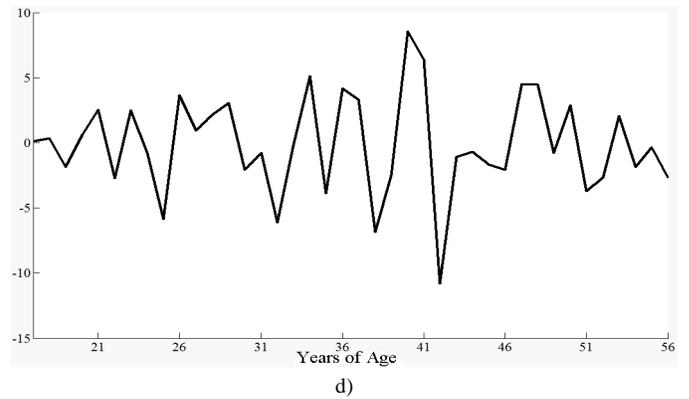

d)

Fig. 2 Quantity signal $\hat{q}$ components: trend (a), periodic ones (b, c), and noise (d).

28.424, 26.547, 24.421, 22.043, 19.646, 17.442, 15.178, 13.1690).

Then, if we substitute $\hat{q}^{(1)}$ from (10) with $\hat{q}^{(1')}$ (and round the sum afterwards) we'll obtain the following modified quantity signal:

$\tilde{q}$ =(2, 73, 127, 129, 133, 129, 117, 124, 92, 120, 92, 105, 87, 89, 79, 80, 83, 93, 78, 85, 79, 61, 62, 63, 59, 30, 38, 28, 24, 16, 21, 16, 10, 11, 4, 3, 7, 4, 2, 3).

Now, let us make some checkup and decompose signal $\tilde{q}$ once again. Appropriate components corresponding to the same index sets we considered before go as follows:

$\tilde{q}^{(1)}$ =(50.931, 90.148, 111.148, 117.605, 118.875, 117.921, 115.819, 113.642, 110.154, 108.694, 105.312, 102.507, 98.915, 95.519, 91.936, 88.599, 85.290, 81.630, 77.128, 72.689, 67.280, 61.954, 57.480, 53.351, 49.103, 44.673, 40.222, 35.740, 31.093, 26.718, 22.547, 18.405, 14.333, 10.343, 6.394, 2.733, −0.165, −2.262, −4.114, −5.392);

$\tilde{q}^{(2)}$ =(−51.014, −11.386, 8.782, 13.694, 13.350, 10.869, 6.351, 2.713, −2.560, −3.293, −6.756, −7.766, −8.906, −8.357, −6.933, −3.961, −0.200, 3.309, 5.220, 7.049, 7.695, 7.534, 6.181, 4.076, 1.523, −1.461, −4.019, −6.101, −7.477, −7.849, −7.105, −5.543, −3.410, −0.990, 1.439, 3.418, 4.907, 6.159, 6.817, 7.331);

$\tilde{q}^{(3)}$ =(4.548, −8.0310, 7.166, −3.244, −0.203, 3.355, −6.395, 8.933, −10.682, 10.648, −9.099, 7.492, −6.036, 4.739, −3.776, 2.592, −2.129, 1.976, −1.419, 0.586, 0.608, −1.469, 2.138, −2.584, 3.016, −3.038, 2.571, −2.137, 1.612, −1.674, 1.509, −1.294, 1.096, −0.867, 0.841, −0.442, −0.227, 1.125, −1.592, 1.4580);

$\tilde{q}^{(4)}$ =(−2.466, 2.269, −0.097, 0.945, 0.977, −3.145, 1.225, −1.288, −4.913, 3.951, 2.543, 2.767, 3.027, −2.901, −2.227, −7.230, 0.039, 6.085, −2.930, 4.676, 3.417, −7.019, −3.799, 8.156, 5.359, −10.174, −0.773, 0.497, −1.229, −1.194, 4.050, 4.431, −2.019, 2.515, −4.674, −2.710, 2.485, −1.022, 0.889, −0.3970).

Appropriate graphs can be found in Fig. 3. The graphical representation of the whole signal can be seen in Fig. 4c.

It is important to emphasize that though signal components have slightly changed, main periodic components' features (like their period, phase and so on) have persisted. This is exactly what we understand by utility preserving.

## IV. CONCLUSIONS AND FURTHER STUDIES

In this paper, we tried to attract attention to some novel techniques in the field of providing anonymity in statistical data, and also attempted to introduce new ways of preserving data utility while altering them in predefined manner.

For the first time, we proposed to apply SSA-Caterpillar to providing group anonymity in population census microfiles, and addressed its main peculiarities. At the same time, we showed that this technique can be efficiently used not only for concealing data distribution features, but also for analyzing them simultaneously. Of course, comprehensive interpreting should be done by qualified sociologists and demographers.

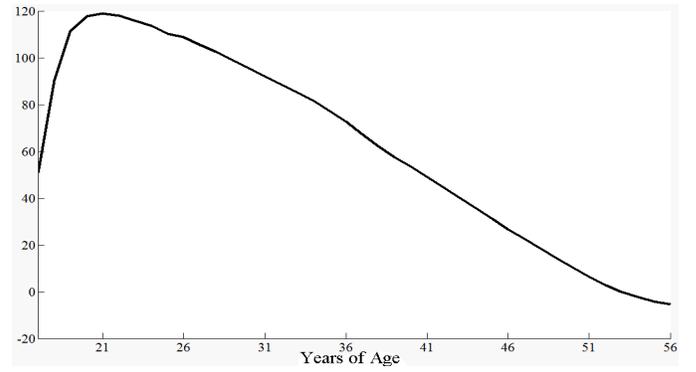

a)

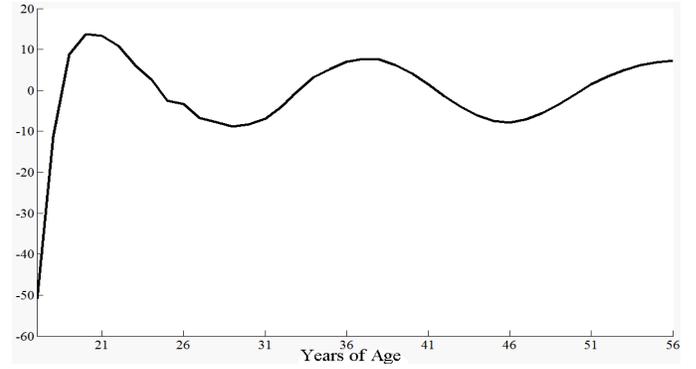

b)

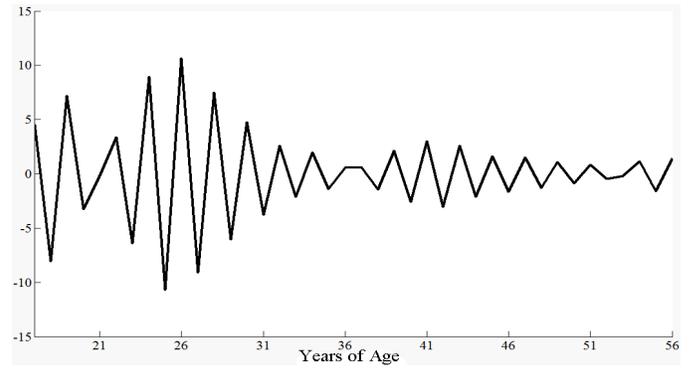

c)

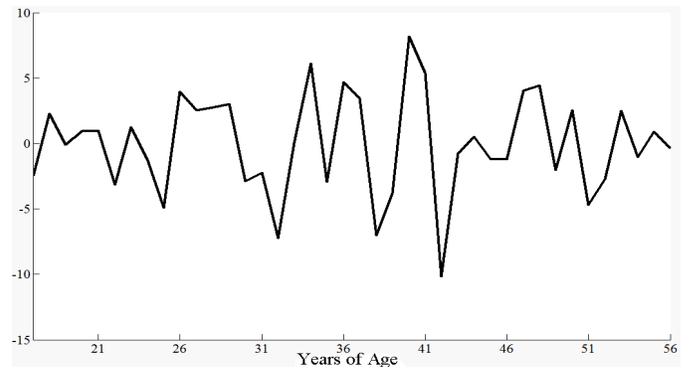

d)

Fig. 3 Quantity signal $\tilde{q}$ components: trend (a), periodic ones (b, c), and noise (d).

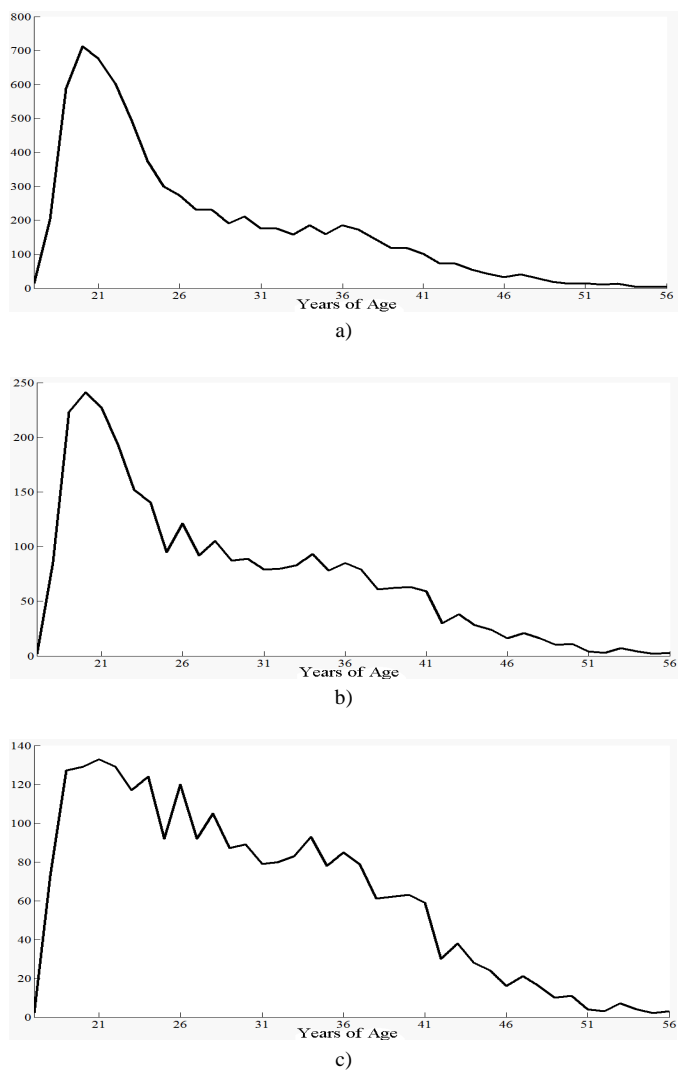

Fig. 4 Quantity signals: initial (a), modified (b), and resultant (c) ones.

And, last but not least, we illustrated the strong importance of correct sampling procedure. Having excluded from consideration an evident cause of incorrect data distribution, we got an opportunity of obtaining meaningful results in the output.

Still, there are some goals to be achieved in the future:

*1) Modifying the Signal*: There has to be developed more formal way of picking up new trend component of the signal.

*2) Data Interpreting*: It would be more helpful to be able to interpret data components from the sociological point of view to define those ones to be preserved in any case, and those ones to be lost without major consequences.

*3) Data Adjusting*: After having altered appropriate goal representations, it is necessary to construct modified microfiles. There needs to be proposed a convenient way of evaluating loss of data utility implied by such constructing procedures.